# Study of the Radiation Hardness of Irradiated AToM Front-End Chips of the BaBar Silicon Vertex Tracker.

S. Bettarini, M. Bondioli, L. Bosisio, G. Calderini[*], S. Dittongo, F. Forti, M. A. Giorgi

*Abstract*--The radiation hardness of the AToM chips of the BaBar Silicon Vertex Tracker has been investigated by means of irradiations with photons from a $^{60}$Co source and 0.9 GeV electrons. The increase in noise and the decrease in gain of the amplifier have been measured as a function of the applied capacitive load and the absorbed dose. Different beam intensities have been used to study the effect of different dose rates to the AToM radiation damage. The chip digital functionalities have been tested up to a dose of 5.5 Mrads for the $^{60}$Co photons and 9 Mrads for the 0.9 GeV electrons. In addition a pedestal shift for the irradiated channels has been observed in the test with electrons but is not present in the irradiation with photons. This effect reproduces qualitatively the behavior observed since 2002 in the front-end electronics of the installed BaBar Silicon Vertex Tracker. After some investigation of the chip layout, this peculiar behavior could be associated to radiation damage in a well-identified component of the AToM. The results of the radiation tests are presented and used to extrapolate the performance and lifetime of the installed detector in the next few years.

## I. INTRODUCTION

THE silicon Vertex Tracker (SVT) of the BaBar experiment [1] at the PEP-II asymmetric B factory [2] is a five layer, double sided, AC coupled silicon microstrip detector. It is a crucial element for the precise measurement of the decay position of B mesons, satisfying the severe constraints imposed by the accelerator design, in terms of geometry of the interaction region and conditions of operation. In exchange for a noticeable delivered luminosity, the integrated dose during the first years of operation is much higher than what expected from design: this is a reason for concern, which recently triggered additional studies on the SVT sensors and front-end electronics to determine the effects of radiation damage up to larger doses. The expected change in performance with radiation, together with an extrapolation of machine background evolution for the next few years is used to evaluate the expected lifetime of the SVT modules.

## II. IRRADIATION TESTS ON THE AToM CHIP

One of the SVT components which are expected to be most sensitive to radiation damage, is represented by the front-end electronics. The AToM ASIC [3] is a radiation hard chip produced by Honeywell in a 0.8 μm CMOS process. The chip is capable of simultaneous signal acquisition, digitization and sparsified readout. Information is shipped out about the hit timestamp and Time-Over-Threshold. The chip is also equipped with a mechanism performing internal charge injection, used for periodic calibration, and a set of diagnostic tools. Radiation damage to the AToM chip has been investigated in order to characterize the increase in noise and decrease in gain with radiation damage, and the presence of possible digital failures. In addition, channel pedestal offset has been studied, since this effect has been observed in the installed system.

The first irradiations of AToM chips on hybrid substrates have been performed at SLAC and LBL using photons from a $^{60}$Co source with doses up to 5 Mrads followed later by a measurement based on 0.9 GeV electrons at the Elettra LINAC (Trieste) up to 9 Mrads, involving also a full inner-layer module with the chips bonded to the sensors. In the case of the run with photons, to irradiate uniformly all the elements, the radiation was thermalized by means of an aluminum-lead box. During all the tests the chips were powered, cooled and provided with an external clock signal. The irradiations were performed in several steps and after each step the analog parameters of the chips (noise, gain, channel pedestal) were measured. No digital failure has been observed in any of the irradiated chips. The noise has been parameterized with two coefficients to model the dependence on the capacitive load: $\sigma = \alpha + \beta C_{load}$. The measured increase in the α and β coefficients was 16%/Mrad and 19%/Mrad respectively, while the gain decreased by about 3%/Mrad (see Fig.1). These values are in very good agreement with the increase in noise and decrease in gain observed in the front end electronics of the installed detector, measured from the calibrations which are performed daily with monitoring puroposes (see Fig.2).

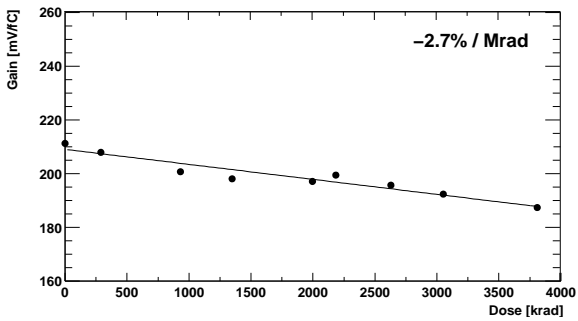

Fig. 1. Gain as a function of the absorbed dose for the irradiated chips. The plot shows a degradation of about 2.7% per Mrad.

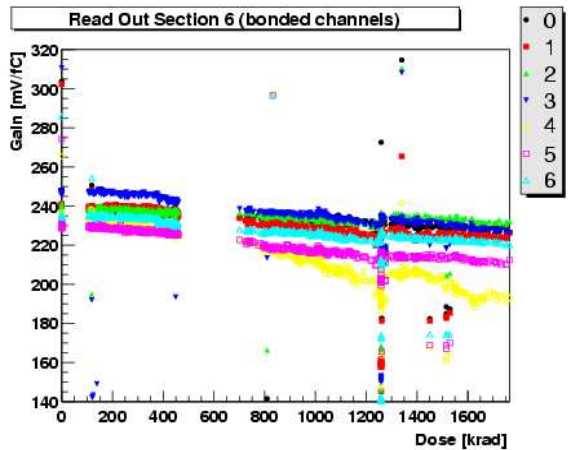

Fig. 2. Gain as a function of the absorbed dose for a module of the installed detector. The different lines correspond to the different chips.

Merging together the information coming from these initial chip irradiations with the results of separate studies on SVT sensors [4], the expected behavior of the signal over noise ratio has been extrapolated as a function of the dose (see Fig.3)

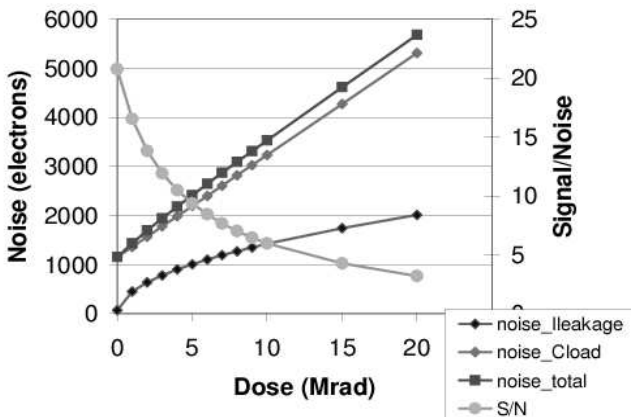

Fig. 3. Predicted noise level (left scale) and signal/noise ratio (right scale) as a function of the integrated dose.

Using a figure of merit of about 10 for the S/N ratio, a reasonable dose budget is set for the SVT at about 5 Mrads. This is not a hard limit but more a scale of the dose at which the degradation of performance starts to become sizeable.

Using background single-beam and collision studies, dose extrapolations can be extracted for the next few years, based on the expected beam currents. These studies indicate that, for modules out of the horizontal plane, the 5 Mrads limit will never be reached over the whole lifetime of the experiment. On the contrary, for modules in the middle plane, the degradation in performance will start to appear after 2006. The interested region is indeed limited to a narrow horizontal strip of 2-3 mm in the central part of the detector.

### III. SHIFT IN CHANNEL PEDESTAL

More recently, the scenario previously described was modified by the observation of a new effect. Since 2002, in the installed system, we are experimenting a change in the pedestal of the channels of the central chips in layer-1 and layer-2 horizontal modules (see Fig.4). The position corresponds to the area most affected by radiation and the interested chips are only on the west side of the detector, the side affected by backgrounds from the High Energy beam.

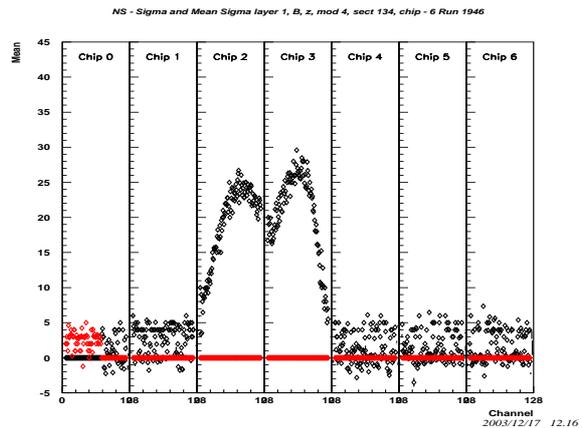

Fig.4. Channel pedestal distribution for a module in the horizontal plane, west side, at a dose of about 1.5 Mrad.

This change was unexpected, since it was not observed during the $^{60}$Co irradiations. The gain and the noise of the chips are not deteriorated, and in principle it is possible to recover the original efficiency by a threshold adjustment. As a matter of fact, this is only partially true. If the pedestal is steadily increasing, the dynamic range of the possible threshold values is soon unusable. In addition, the pedestal distribution over the channels of a chip is rather wide, while a single threshold can be set for the whole chip. This makes the efficiency optimization very difficult. A detailed analysis shows that, when exposed to radiation, after an initial increase the pedestal of each channel reaches a sort of plateau and then starts to decrease towards normal values (see Fig.5). Even if in practice this behavior seems to suggest that the short-term operation of the detector is not in danger, the long-term evolution needs to be predicted and the causes of the process need to be understood better.

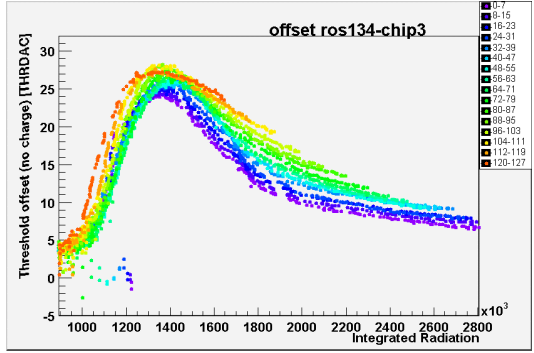

Fig. 5. Pedestal evolution as a function of the dose for groups of channels of one of the affected chips. The pedestal increases after a dose of about 1 Mrad, then reaches a saturation and drops back towards normal values. Different groups of channels behave in a very similar way, even if they reach the corresponding doses at different times.

Since the effect was not seen with photons and since the main background component at PEP-II is known to be from electromagnetic cascade, we decided to study the chip behavior during irradiation with electrons in the GeV range. The program of measurements was made using the Elettra LINAC (Trieste) in 2004. Chips loaded on a hybrid and on a full inner layer module have been irradiated. An x-y stage allowed to move the sample across the beam, while a DAQ system, providing power and clock, was used to run digital test during the irradiation and to provide calibration runs between irradiation steps (see Fig. 6). No permanent digital failure has been observed up to a dose of 9 Mrad.

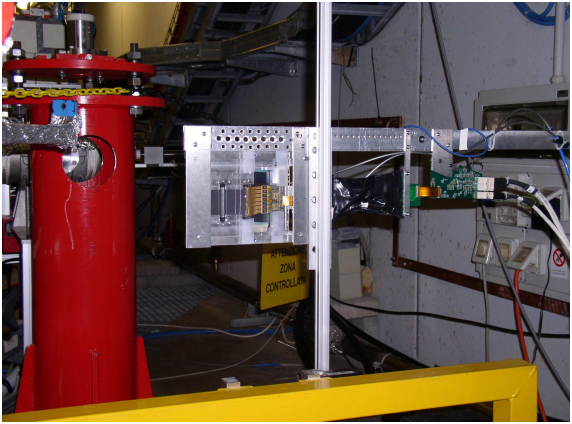

Fig. 6. Irradiation setup at Elettra. The high density interconnect and the inner layer module are moved across the beam by a xy-stage and are connected to a DAQ system, allowing calibrations between irradiation steps.

There are seven chips on each side ($\phi$,z) of the HDI and module, with numbers assigned from 0 to 6. A preliminary run was made on a single chip (chip #2) with the goal of optimizing the LINAC beam current. A compromise was found between the need to reach a 10 Mrads dose in a reasonable time and the requirement not to damage the chips due to the instantaneous dose. The electron peak fluence rate selected for the irradiation was $\phi = 3.75 \times 10^{11}$ e-/cm$^2$s, corresponding to a peak dose rate of about 10 Krad/s. The beam size was $\sigma_{beam}$= 1.8mm.

Using the selected rate, in addition to the chip already irradiated, a scan of three more chips loaded on the High Density Interconnect (Chip #3,4,5) was performed. This allowed to reach an integrated dose of about 9 Mrads on chip #2 and about 3.2 Mrads on the others. The channel pedestal distribution changed during the irradiation, showing the expected drift, at least in a qualitative way (see Fig.7)

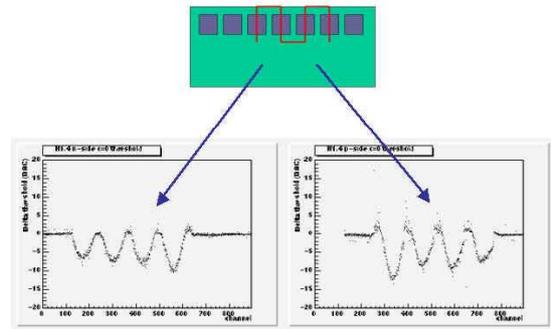

Fig. 7. Pedestal change after 3.2 Mrads for the irradiated chips. The beam path across the four irradiated chips is sketched in the upper figure. There are 128 channels for each chip. The profile of the beam scan is clearly visible in the plot showing the pedestal change after the irradiation. The left and right plots are the chips on the n-side and p-side of the High Density Interconnect.

The evolution of pedestal is shown in Fig.8. The drift is rather fast at the beginning, then the pedestal reaches a plateau. When the irradiation is suspended there is some annealing effect, after which the drift continues when the irradiation is resumed. Again, saturation is achieved.

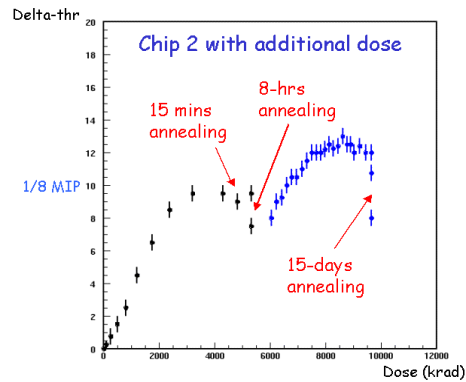

Fig. 8. Pedestal change as a function of dose. After an initial increase, the pedestal reaches a plateau. During the night stop, the pedestal shows some annealing effect (center part of the plot). When irradiation is resumed, the pedestal reaches a saturation. The last point was taken after a 15 days-annealing

After the HDI, also the chips loaded on a inner layer module were irradiated up to 6 Mrads. The results are very similar to what obtained on the HDI (see Fig.9)

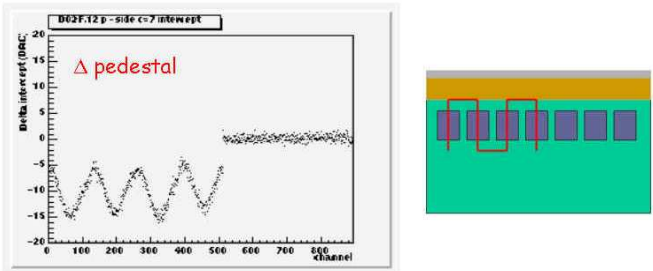

Fig. 9. Pedestal change after 6 Mrads for the irradiated chips on the module. The beam path across the four irradiated chips is sketched in the figure. Again, the profile of the beam scan is clearly visible in terms of pedestal variation.

No gain or noise variation has been observed in the calibrations, and this has been confirmed by measurements taken after the irradiation, generating charge inside the silicon by a 1060nm LED.

The pedestal variation as a function of the dose (see Fig.10) shows a distribution similar to what measured in the case of the HDI. After an initial variation, the pedestal approaches saturation. A clear annealing effect is visible at the end of the irradiation.

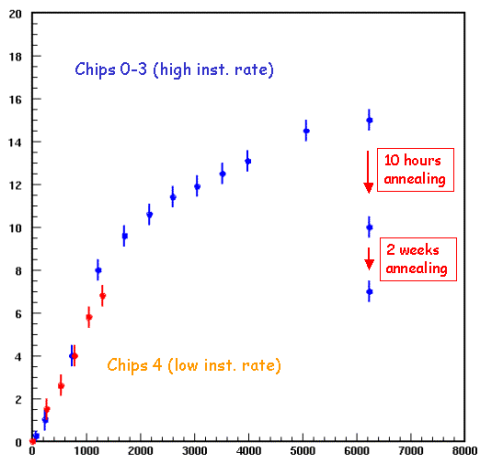

Fig. 10. Pedestal change as a function of dose for the chips of the inner layer module. Also here, after an initial increase, the pedestal reaches a saturation. At the end of the irradiation there is a clear annealing effect. The last point was taken 15 days after the end of the measurements.

One more chip (Chip #4) has been irradiated at a lower rate, to investigate a possible dependence of the effect on the dose rate. No difference was seen with respect to the other chips.

## IV. COMPARISON WITH THE INSTALLED SYSTEM AND INTERPRETATION OF RESULTS

The behavior shown by the chips irradiated at Elettra is presenting a number of common features to what observed in the real experiment. There is indeed some quantitative difference. The maximum pedestal deviation during the test was of the order of about 15 counts, corresponding to about 1/6 MIP. In the real system, the saturation was reached at about 25 counts (see Fig.5). The difference can be attributed to the necessarily fast rate at which the dose has been delivered during the test, which is orders of magnitude larger with respect to the PEP-II interaction region. Some instantaneous damage saturation is likely to be the cause of the observed quantitative discrepancy. On the contrary, in the irradiation, the residual pedestal shift after the annealing effect is about 5-8 counts, and the same change is present in the real experiment, after the pedestal goes down towards normal values. In spite of the numerical discrepancies, the main achievement of the test has been to determine that the observed shift in pedestal can be obtained with an electromagnetic cascade, the same nature of the main background component present in the PEP-II interaction region. The irradiation, pushed to 9 Mrads, gave indication that the evolution of the shift is not expected to represent a danger for the SVT operation even in the long-term perspective.

Even if the shift of the threshold voltage with radiation is a known effect for sample CMOS structures, special attention had been put in the design and layout of the AToM chip to prevent this problem. In particular, extensive use of mirrored components had been made, to mitigate the effects of temperature changes and radiation damage. Given the result of the test beam, the schematics of the AToM chip have been investigated to search for a possible weakness in the design. The analysis has shown the presence of two possible weak points, one in the shaper and one in the comparator, where the principle has not been respected. Detailed simulations are underway to study the effect in more detail.

## V. CONCLUSIONS

We have characterized the radiation hardness of the BaBar SVT AToM front-end chips to extrapolate the lifetime and the expected reduction in performance. We used $^{60}$Co sources and electrons in the GeV range. A radiation budget of 5 Mrads has been set; this dose will never be reached by the SVT modules out of the horizontal plane, while will be reached after 2006 by the central part of the modules in the beam bending plane. In addition, an unexpected effect has been observed in the installed detector, consisting in a drift of the pedestal of the channels more exposed to radiation. After an initial increase the pedestal reaches a level of saturation and then decreases down towards normal values. We were able to reproduce the drift after irradiation with 0.9 GeV electrons and we have indication that we understood the cause. During the test an integrated dose of about 9 Mrads was reached in order to study the long-term evolution of the effect and to exclude that the behavior could show a critical trend for the system operation in the future.